\documentstyle[12pt]{article}
\begin{document}

\topmargin=-1.5cm
\evensidemargin=0cm
\oddsidemargin=0cm
\date{}

\newcommand{\BQ}{\begin{equation}}
\newcommand{\EQ}{\end{equation}}
\newcommand{\BQA}{\begin{eqnarray}}
\newcommand{\EQA}{\end{eqnarray}}
\newcommand{\half}{\frac{1}{2}}
\newcommand{\NN}{\nonumber \\}
\newcommand{\REN}{{\rm ren}}
\newcommand{\E}{{\rm e}}
\newcommand{{\BPsi}}{\bar{\Psi}}
\newcommand{{\bpsi}}{\bar{\psi}}
\newcommand{\GF}{\gamma_{5}}
\newcommand{\Gmu}{\gamma^{\mu}}
\newcommand{\Gnu}{\gamma^{\nu}}
\newcommand{\gmu}{\gamma_{\mu}}
\newcommand{\gnu}{\gamma_{\nu}}
\newcommand{\GSLP}{\frac{g^{2}\lambda}{\pi}}
\newcommand{\GLP}{\frac{g\lambda}{\pi}}
\newcommand{\GSLSP}{g^{2}\lambda/\pi}
\newcommand{\GLSP}{g\lambda/\pi}

\renewcommand{\thefootnote}{\fnsymbol{footnote}}

\begin{flushright}
UT-Komaba\\
93-11\\
July, 1993
\end{flushright}

\vspace*{2cm}

\begin{center}
{\large {\bf Hamiltonian Formulation of the Smooth Bosonization\\
and Local Gauge Symmetry of the Massless Thirring Model}}

\vspace{1.5cm}

{\large T. Ikehashi}

\vspace{1.5cm}

{\it Institute of Physics, University of Tokyo, \\ Komaba, Meguro-Ku,
Tokyo 153, Japan} \end{center}

\vspace{2.5cm}

\noindent {\bf Abstract}:
Using the hamiltonian formalism, we investigate the smooth bosonization method
in which bosonization and fermionization are carried out through a specific
gauge-fixing
of an enlarged gauge invariant theory. The generator of the
local gauge symmetry, which cannot be derived from the lagrangian of the
enlarged theory, is obtained by making a canonical transformation.
We also show that
the massless Thirring model possesses a similar local gauge symmetry for a
speific value of the coupling constant.

\vspace{1cm}

\newpage

 A few years ago, Alfaro and Damgaard \cite{AD} developed an idea that
field-enlarging
transformations can be used to find equivalent formulations of the same theory.
Recently, Damgaard, Nielsen and Sollacher \cite{DNS} have applied this idea
to the 1+1
dimensional abelian bosonization. Their procedure consists of two steps:
First, they introduced a bosonic field to a fermionic theory through a chiral
transformation. The theory thus enlarged possesses a local gauge symmetry.
Second, they showed that a particular gauge-fixing condition enables us to
interpolate smoothly between fermionic and bosonic formulations.
The accomplishment of this program seems to rely deeply on the use of the
path-integral method since the chiral Jacobian of the fermionic measure plays
an important role in forming the local gauge symmetry.
Another reason which makes one to prefer the path-integral method is that
the generator of the local gauge symmetry, which
is crucial in the hamiltonian formalism, cannot be found from the enlarged
lagrangian. In this letter, we shall show that it is nevertheless
possible to formulate the scheme of Damgaard et.\ al. in the hamiltonian
formalism.

 First, we would like to clarify the correspondent of the Jacobian in the
hamiltonian formulation. We start from a quantized free Dirac fermion with
the lagrangian density\footnote[2]{Our conventions are $g_{\mu\nu}={\rm diag}
(1,-1),\GF =\gamma^0\gamma^1,\Gmu \GF=\epsilon^{\mu\nu}\gamma_{\nu},
\epsilon_{01}=-\epsilon^{01}=1$.}
\BQ
{\cal L}=i\BPsi \Gmu \partial_{\mu}\Psi .
\EQ
Let us suppose that the axial vector current is so regularized that
under a local chiral transformation $\Psi (x)\rightarrow \E ^{-ig\gamma_{5}
\phi (x)}\Psi (x), \BPsi (x)\rightarrow \BPsi (x)\E ^{-ig\GF \phi (x)}$
($g$ being a coupling constant), it transforms as
\BQ
J_{5}^{\mu}=\BPsi \Gmu \GF \Psi \rightarrow J_{5}^{\mu}+
\GLP \partial^{\mu}\phi ,
\EQ
where a parameter $\lambda$ represents an ambiguity of the regularization
\cite{HKT,Bnj}. Now let us apply the chiral transformation to
the lagrangian by means of iterating infinitesimal chiral rotations. Then, the
first infinitesimal transformation yields
\BQ
{\cal L}\rightarrow i\BPsi \Gmu \partial_{\mu}\Psi +g\BPsi \Gmu \GF \Psi
\partial_{\mu}(\delta \phi). \label{firstL}
\EQ
Noting that the axial vector current in (\ref{firstL})
is also transformed by further infinitesimal transformations, we obtain the
transformed lagrangian
\BQ
{\cal L}=i\BPsi \Gmu \partial_{\mu}\Psi +g\BPsi \Gmu \GF \Psi \partial_{\mu}
\phi+\half \GSLP \partial_{\mu}\phi\partial^{\mu}\phi. \label{enL}
\EQ
The present procedure just parallels what Roskies and Schaposnik
\cite{RosSch} showed in the path-integral method. The last term of (\ref{enL})
corresponds to the
contribution of the Jacobian. This shows that, in the hamiltonian formalism,
the change of
the regularized current represents the effect of the Jacobian. The explicit
form of the regularization will be presented soon.

 In the path-integral formulation, the
function $\phi (x)$ can be promoted to a dynamical variable by integrating
over $\phi$. As a result, the enlarged lagrangian (\ref{enL}) exhibits a
local gauge symmetry
\BQA
&&\phi (x)\rightarrow \phi (x)-\alpha (x),\,\,\Psi (x)\rightarrow \E ^{-ig\GF
\alpha (x)}\Psi (x),\,\,\BPsi (x)\rightarrow \BPsi (x)\E ^{-ig\GF \alpha (x)},
\label{LGS}\\
&&J_{5}^{\mu}(x)\rightarrow J_{5}^{\mu}(x)+\GLP \partial^{\mu}\alpha (x).
\label{LGSJ5}
\EQA
In the hamiltonian formulation, however, it is not clear how to promote
$\phi$ to
be dynamical at this stage since $\Psi$ is $already$ quantized and is a
free Dirac fermion. To circumvent this difficulty, we decide to start from
a classical theory defined by a lagrangian density
\BQ
{\cal L}=i\BPsi \Gmu \partial_{\mu}\Psi +g\BPsi \Gmu \GF \Psi \partial_{\mu}
\phi+\half \partial_{\mu}\phi\partial^{\mu}\phi. \label{cL}
\EQ
By comparison with (\ref{enL}), it is expected that the quantized system
exhibits a local gauge symmetry (\ref{LGS}) and (\ref{LGSJ5}) when currents
are regularized to
satisfy (\ref{LGSJ5}) and when $\GSLSP =1$. The lagrangian density (\ref{cL})
with the
mass term $\half m^{2}\phi^{2}$ is known as the RS model
\cite{Bnj,RS}
and is solved for
$\GSLSP <1$. (We ignore the infrared divergence of the massless free boson
since we are interested in boson-fermion correspondence at the operator level.)
The classical
equation of motion has a solution
$\Psi (x)=\E ^{ig\GF \phi (x)}\psi (x)$,
where $\psi (x)$ is a free fermion ($i\Gmu \partial_{\mu}\psi=0$) and
$\phi (x)$ is a harmonic field ($\Box \phi =0$). The canonical momenta
conjugate to $\phi$ is given by
\BQ
\Pi=g\BPsi \gamma ^{0}\GF\Psi +\partial^{0}\phi.
\EQ
Quantization is performed by imposing the equal-time canonical commutation
relations;
\BQA
&&[\Pi(x),\phi(y)]=-i\delta(x^{1}-y^{1}),\,\,\{ \Psi_{\alpha}(x),\BPsi_{\beta}
(y) \}=(\gamma_{0})_{\alpha \beta}\delta(x^{1}-y^{1}),\NN
&&[\phi(x),\phi(y)]=[\phi(x),\Psi(y)]=\{ \Psi_{\alpha}(x),\Psi_{\beta}
(y) \}=0.
\EQA
We define the regularization of current as \cite{Bnj}
\BQ
J^{\mu}(x)=\lim_{\varepsilon\rightarrow0}\BPsi(x+\varepsilon )\Gmu \exp\left[
ig\int_{x}^{x+\varepsilon}dy_{\rho}\{\lambda \epsilon^{\rho \nu}\partial_{\nu}
\phi+(\lambda+1)\GF \partial^{\rho}\phi\} \right] \Psi(x)-{\rm v.e.v.},
\EQ
where v.e.v. stands for the ``vacuum expectation value". The vector and axial
vector
currents thus regularized can be rewritten in terms of free fermion currents
$j^{\mu}=:\bpsi \Gmu \psi:$, $j_{5}^{\mu}=:\bpsi \Gmu \GF \psi:$ and free
boson $\phi$ by inserting the normal ordered solution;
\BQA
J^{\mu}&=&j^{\mu}-\GLP \epsilon^{\mu\nu}\partial_{\nu}\phi, \label{J}\\
J_{5}^{\mu}&=&j_{5}^{\mu}-\GLP \partial^{\mu}\phi. \label{J5}
\EQA
Using (\ref{J5}), one can show that the classical equation of moton for
$\phi$ is modified to $(1-\GSLSP )\Box \phi =0$ which disappears at
$\GSLSP =1$.
It is also
worth noting that according to the analysis of the RS model, there is a
wavefunction and coupling constant renormalization
$\phi=Z^{1/2}\phi^{(\REN )}$, $g=Z^{-1/2}g^{(\REN )}$ with
$Z=(1-\GSLSP )^{-1}$.
These results indicate that the
model becomes singular at the critical value $\GSLSP =1$. However, the
emergence of the local gauge symmetry is not clear from these considerations
since there are no first-class constraints (=generator of the local gauge
symmetry). To see how the local gauge symmetry appears, it is useful to
perform a canonical transformation
\BQ
\Psi(x)=:\E ^{ig\GF \phi(x)}:\psi(x) \label{can}
\EQ
inspired by the form of the exact solution. The normal ordering is not needed
at $\GSLSP =1$, since the equation of motion for $\phi$ vanishes at this value.
By this transformation, the
lagrangian density is rewritten as
\BQ
{\cal L}=i\bpsi \Gmu \partial_{\mu}\psi
+\half (1-\GSLP )\partial_{\mu}\phi\partial^{\mu}\phi. \label{canL}
\EQ
Note that insertion of (\ref{can}) into the lagrangian should be done
$gradually$ as in the derivation of (\ref{enL}). The canonical momenta of
$\phi$ defined from this lagrangian density is
\BQ
\pi^{\phi}=(1-\GSLP )\partial^{0}\phi. \label{defpi}
\EQ
The canonical variables are now $(\bpsi ,\psi,\pi^{\phi},\phi)$ with the
canonical commutation relations;
\BQA
&&[\pi^{\phi}(x),\phi(y)]=-i\delta(x^{1}-y^{1}),\,\,\{ \psi_{\alpha}(x),
\bpsi_{\beta}
(y) \}=(\gamma_{0})_{\alpha \beta}\delta(x^{1}-y^{1}),\NN
&&[\phi(x),\phi(y)]=[\phi(x),\psi(y)]=\{ \psi_{\alpha}(x),\psi_{\beta}
(y) \}=0. \label{com2}
\EQA
The definition (\ref{defpi}) implies that a first-class constraint $\Phi_{1}
\equiv \pi^{\phi}\approx 0$ emerges when $\GSLSP =1$. This is indeed the
generator of the local gauge symmetry (\ref{LGS}) and (\ref{LGSJ5}) as can be
verified through (\ref{J5}), (\ref{can}) and (\ref{com2}).

 To understand the situation at $\GSLSP =1$, it may be intuitive to consider
a general case. Let us consider a lagrangian density
${\cal L}(\phi, \partial_{\mu}\phi)$. If we introduce a field $\theta$
by a canonical transformation
\BQ
\phi=f(\phi',\theta), \label{canex}
\EQ
the Hilbert space gets enlarged and acquires a local gauge symmetry with
a first-class
constraint \cite{AD,HK}
\BQ
\Phi_{1}'\equiv \pi'-\Pi'\left( \frac{\partial f}{\partial \phi'}
\right)^{-1}\frac{\partial f}{\partial \theta}\approx 0, \label{Phid}
\EQ
where $\pi'=\partial {\cal L}(f, \partial_{\mu}f)/\partial \dot{\theta}$ and
$\Pi'=\partial {\cal L}(f, \partial_{\mu}f)/\partial \dot{\phi'}$. In the
enlarged Hilbert space constructed from ($\Pi',\phi',\pi',\theta$), the
physical state $|{\rm phys}'\rangle$ should satisfy
$\Phi_{1}'|{\rm phys}'\rangle=0$ owing to the gauge invariance.
On the other
hand, if we solve $\phi'$ from (\ref{canex}) as $\phi'=g(\phi,\theta)$ and
consider this as a canonical transformation for the enlarged Hilbert space,
we get a first-class constraint
\BQ
\Phi_{1}\equiv \pi\approx 0,
\EQ
where $\pi=\partial {\cal L}(f, \partial_{\mu}f)/\partial \dot{\theta}$ with
$f=f(g(\phi,\theta),\theta)$. This constraint has appeared simply because
our original lagrangian density ${\cal L}(\phi, \partial_{\mu}\phi)$ did
not contain $\theta$. The physical state for the transformed Hilbert space
$(\Pi,\phi,\pi,\theta)$
($\Pi=\partial {\cal L}(\phi, \partial_{\mu}\phi)/\partial \dot{\phi}$)
should satisfy $\Phi_{1}|{\rm phys}\rangle=-i\delta/\delta \theta
|{\rm phys}\rangle=0$. This means that only $(\Pi,\phi)$ are physical and
$(\pi,\theta)$ are redundant degrees of freedom.
It is important to recognize here
that we can investigate gauge-transformation properties of
$(\Pi',\phi',\pi',\theta)$ in terms of $\Phi_{1}$,
because
$\Pi',\phi'$ and $\pi'$ can be solved with respect to $\Pi,\phi,\pi$ and
$\theta$.
As for the present case,
we could not find the first-class constraint of the form (\ref{Phid}) but we
can nevertheless know the transformation laws for the operators of
$(\BPsi ,\Psi,\Pi,\phi)$ using $\Phi_{1}$.

 We can also examine the gauge-fixing condition of Damgaard et.\ al. with
the help of
$\Phi_{1}=\pi^{\phi}\approx0$. The
physical state should satisfy $\Phi_{1}|{\rm phys}\rangle=-i\delta/\delta
\phi|{\rm phys}\rangle=0$. In the language of
$(\bpsi ,\psi,\pi^{\phi},\phi)$, it implies that the physical state should be
constructed from $(\bpsi ,\psi)$. One can translate this into the language of
$(\BPsi ,\Psi,\Pi,\phi)$ by the use of (\ref{J}), (\ref{J5}) and (\ref{can}).
For example, a combination $J_{5}^{\mu}+\GLSP \partial^{\mu}\phi$ is observable
in $(\BPsi ,\Psi,\Pi,\phi)$ since this is equivalent to $j_{5}^{\mu}$ which
belongs to $(\bpsi ,\psi)$. Damgaard et.\ al. used this combination as a
guide to
gauge-fixing. Their gauge-fixing condition has a form
\BQA
\Phi_{2}(x)&=&\int_{x_{0}}^{x}d\xi_{\mu}\left\{ \Delta(J_{5}^{\mu}+\GLP
\partial^{\mu}\phi)-\GLP \partial^{\mu}\phi\right\} \NN
&=&\int_{x_{0}}^{x}d\xi_{\mu}\left( \Delta j_{5}^{\mu}-\GLP
\partial^{\mu}\phi\right) \approx0,
\EQA
where $\Delta$ is a parameter. The path-independence of this expression is
evident if we recall that the axial vector current for free fermion can be
written in a form $j_{5}^{\mu}=\partial^{\mu}\varphi$
\cite{KAAR}. Therefore
when $\Delta=1$, we get $\partial^{\nu}_{x}\Phi_{2}=J_{5}^{\nu}(x)\approx0$
which indicate that the fermionic current cannot be observed. On the other
hand,
$\Delta=0$ implies that the bosonic field $\phi$ vanishes from the physical
states. This can be verified in terms of the Dirac brackets \cite{Drc}.
Let us employ a space-like path
\BQ
\Phi_{2}(x)=\Delta\int_{-\infty}^{x^{1}}d\xi_{1}j_{5}^{1}(x^{0},\xi^{1})
-\GLP \{ \phi(x^{0},x^{1})-\phi(x^{0},-\infty)\} \approx0.
\EQ
Then, observing that $[\Phi_{i}(x),\Phi_{j}(y)]=i(\GLSP )\epsilon_{ij}
\delta(x^1-y^1)$
($\epsilon_{12}=-\epsilon_{21}=1$), we obtain the Dirac bracket for arbitrary
operators $A$ and $B$,
\BQ
[A(x),B(y)]_{D}=[A(x),B(y)]+\frac{i\pi}{g\lambda}\int dz^1
[A(x),\Phi_{i}(z^1)]\epsilon_{ij}[\Phi_{j}(z^1),B(y)].
\EQ
Noting that $\Pi=gj_{5}^{0}+\pi^{\phi}$ and using
$[j^{0}(x),j^{1}(y)]=(i/\pi)\partial_{x}^{1}\delta(x^1-y^1)$,
the Dirac brackets for $\Pi,\phi$ and
fermionic currents lead to
\newcommand{\JJ}{[J_{5}^{0}(x),J_{5}^{1}(y)]_{D}}
\BQA
[\Pi(x),\phi(y)]_{D}&=&-i\frac{\Delta}{\lambda}\delta(x^{1}-y^{1}),\\
\JJ&=&(1-\Delta)\frac{i}{\pi}\partial^{1}_{x}
\delta(x^1-y^1).
\EQA
The commutators for $\Pi,\phi$ and fermionic currents vanish for $\Delta=0$
and $\Delta=1$, accordingly. Since the non-existence of the commutators
can be interpreted as disappearance of these variables from the physical
space,
these results support the
observation of Damgaard et.\ al. It is also tempting to calculate
the anti-commutation relation for fermionic fields.
With the help of
\BQ
[\left( \E ^{ig\GF \phi(x)}\right)_{\alpha\beta},\psi_{\gamma}(y)]_{D}=
\Delta\frac{i\pi}{\lambda}\left( \GF \E ^{ig\GF \phi(x)}\right)_{\alpha\beta}
\psi_{\gamma}(y)\theta(x^1-y^1),
\EQ
we find
\BQA
\{ \Psi_{\alpha}(x),\BPsi_{\beta}(y) \}&=&(\gamma_{0})_{\alpha \beta}
\delta(x^{1}-y^{1})\NN
&+&\Delta\frac{i\pi}{\lambda}\left[ \theta(y^1-x^1)
\left( \E ^{ig\GF \phi(x)}\right)_{\alpha\gamma}\left(\BPsi (y)\GF
\right)_{\beta}\left(\E ^{-ig\GF \phi(x)}\Psi (x)\right)_{\gamma}
\right.\NN
&&-\left.\theta(x^1-y^1)\left(\BPsi (y)\E ^{-ig\GF \phi(y)}\right)_{\gamma}
\left(\GF \Psi (x)\right)_{\alpha}\left( \E ^{ig\GF \phi(y)}
\right)_{\gamma\beta}\right].
\EQA
The anti-commutation relation reduces to the usual one for $\Delta=0$,
but it does not vanish at $\Delta=1$ which is contrary to naive expectation.
This suggests that while the smooth bosonization method is successful in
deriving the bosonization relations for currents (and the mass-term relations
which we do not discuss in this letter), the complete elimination of the
fermionic field is not possible. In the path-integral method, this may
correspond to the fact that one
cannot decouple the generating functional into fermionic and bosonic part when
there is a source term of a form $\BPsi \eta+\bar{\eta}\Psi$.

 Finally, we would like to point out the similarity between the massless RS
model (\ref{cL}) and the massless Thirring model. The lagrangian density of
the massless Thirring model is given by
\BQA
{\cal L}&=&i\BPsi \Gmu \partial_{\mu}\Psi+\frac{g}{2}(\BPsi \Gmu \Psi)^2\NN
&=&i\BPsi\Gmu \partial_{\mu}\Psi+\BPsi \Gmu \Psi A_{\mu}-\frac{1}{2g}A_{\mu}
A^{\mu},
\EQA
where we have introduced an auxiliary field $A_{\mu}$ in the second line.
Parametrizing the auxiliary field as $A_{\mu}=\epsilon_{\mu\nu}\partial^{\nu}
\phi+\partial_{\mu}\eta$, we get
\BQ
{\cal L}=i\BPsi \Gmu \partial_{\mu}\Psi+\BPsi \Gmu \Psi(\epsilon_{\mu\nu}
\partial^{\nu}\phi+\partial_{\mu}\eta)+\frac{1}{2g}(\partial_{\mu}\phi
\partial^{\mu}\phi-\partial_{\mu}\eta\partial^{\mu}\eta).
\EQ
As can be seen from this expression, this is almost the lagrangian density of
the massless RS model except for the existence of the indefinite
metric field $\eta$
and trivial rescaling of $g$ and $\phi$. Indeed, as in the case of the massless
RS model, the massless Thirring model exhibits a local gauge symmetry for a
specific value of the coupling constant $g$ (and a parameter $\lambda$
representing regularization ambiguity).
Instead of repeating arguments in the hamiltonian formalism, let us
employ the path-integral method this time. Making a change of variable
$\Psi=\E^{-i\GF \phi +i\eta}\psi$ in the partition function and taking into
account the change of the fermionic measure \cite{Bnj};
\BQ
{\cal D}\Psi{\cal D}\BPsi ={\cal D}\psi{\cal D}\bpsi \exp[-
\frac{i\lambda}{2\pi}\int d^{2}x\partial_{\mu}\phi\partial^{\mu}\phi],
\EQ
we get
\BQA
Z&=&\int{\cal D}\Psi{\cal D}\BPsi {\cal D}\phi {\cal D}\eta
\exp\left[i\int d^{2}x\left\{ i\BPsi \Gmu \partial_{\mu}\Psi+\BPsi \Gmu \Psi
(\epsilon_{\mu\nu}
\partial^{\nu}\phi+\partial_{\mu}\eta)\right.\right. \NN
&&+\left.\frac{1}{2g}\left.(\partial_{\mu}\phi
\partial^{\mu}\phi-\partial_{\mu}\eta\partial^{\mu}\eta)\right\} \right]\NN
&=&\int{\cal D}\psi{\cal D}\bpsi {\cal D}\phi {\cal D}\eta
\exp\left[i\int d^{2}x\{i\bpsi \Gmu \partial_{\mu}\psi+
\half\left(\frac{1}{g}-\frac{\lambda}{\pi}\right)\partial_{\mu}\phi
\partial^{\mu}\phi
-\frac{1}{2g}\partial_{\mu}\eta\partial^{\mu}\eta\} \right].\NN \label{Thrg}
\EQA
When $g\lambda/\pi=1$, the integration over $\phi$ diverges.
This implies the emergence of a local gauge symmetry in the path-integral
formulation. The generator of this local gauge symmetry will be
$\Phi_{1}\equiv\pi^{\phi}\approx 0$, where $\pi^{\phi}$ is the conjugate
momenta of $\phi$ defined from the lagrangian in the last line of (\ref{Thrg}).
Exploiting the gauge-fixing condition $\Phi_{2}\equiv\phi\approx 0$,
we may eliminate the redundant degrees of freedom $(\pi^{\phi},\phi)$,
and the remaining physical part describes free fermion.

 In conclusion, we have showed that a local gauge symmetry which includes
the contribution from the fermionic Jacobian can also be treated in the
hamiltonian formalism. The generator obtained from the canonical-transformed
lagrangian enabled us to use the Dirac method for constrained systems.
We have reproduced the results of Damgaard et.\ al. for bosonization relations
of currents. It may be interesting to apply this method to the non-abelian
bosonization \cite{Witten}.

\vspace{1cm}

\noindent {\bf Acknowledgements}:
 I would like to thank K.\ Ohta, K.\ Itakura, I.\ Ichinose and H.\ Mukaida
for discussions. I am greatful to K.\ Ohta for his careful reading of the
manuscript.

\end{document}